\documentstyle[preprint,aps]{revtex}
\begin{document}
\draft

\title{Coulomb Drag in Double Layers with Correlated
Disorder}
\author{I.V. Gornyi$^{1,2}$, A.G.
Yashenkin$^{1,3}$, and  D.V. Khveshchenko$^1$}
\address{
$^1$ NORDITA, Blegdamsvej 17, DK-2100, Copenhagen, Denmark\\
$^2$ A.F.Ioffe Physical-Technical Institute, Polytechnicheskaya 26,
St.Petersburg 194021, Russia\\
$^3$ Petersburg Nuclear Physics
Institute, Gatchina, St. Petersburg 188350, Russia}
\date{\today}
\maketitle
\tightenlines

\begin{abstract}
We study the effect of correlations
between impurity potentials in different layers on the Coulomb drag
in a double-layer electron system. It is found that for strongly
correlated potentials the drag in the diffusive
regime is considerably enhanced as compared to conventional predictions.
The appropriate experimental conditions are discussed, and the new
experiments are suggested.

\end{abstract}





Over the past decade the frictional drag in
double-layer two-dimensional electron systems has been a
subject of extensive experimental \cite{exp} and
theoretical \cite{convth,CF,other,tun}
studies. This phenomenon is manifested in the appearance
of current $I_2$ or voltage $V_2$ in the ``passive'' layer
2 when the applied voltage $V_1$ causes the current $I_1$
to flow in the ``active'' layer 1. The strength of the drag
is characterized by either transconductivity
$\sigma_{21}= (I_2 / V_1 )_{V_2 =0}$ or transresistivity
$\rho_{21} = (V_2 / I_1 )_{I_2 =0}$, which are related one to another
as $\rho_{21} = - \sigma_{21} (\sigma_{11} \sigma_{22} -
\sigma_{12} \sigma_{21})^{-1} \approx - \sigma_{21}
\sigma_{11}^{-2}$ where $\sigma_{ii}$ are the intrinsic
conductivities of the layers.

In the absence of tunneling (cf.\ Ref.\ \cite{tun}), the
drag arises due to interlayer momentum transfer mediated by
inelastic scattering (mainly, Coulombic) of carriers
that belong to different layers.
Hence, the drag provides
a convenient tool for studying electron correlations in the
coupled two-dimensional mesoscopic systems, including such
important characteristics as the electron polarization
function and the screened interlayer interaction.

In the conventional theory \cite{convth}, the carriers
in each layer are scattered by their own impurity potentials.
As a result, the processes contributing to $\sigma_{21}$ can be
understood in terms of coupling between {\it independent} thermal
density fluctuations in different layers. The phase space 
available to the thermal excitations is small and limited by
energies $\omega \lesssim T$. Therefore, the drag effect rapidly
vanishes with decreasing temperature. For instance,
$\rho_{21} \propto T^2$ ($T^2 \ln T$) in a clean
(dirty) normal metal \cite{convth} and $\rho_{21}
\propto T^{4/3}$ for composite fermions in double-layers of
electrons in the half-filled Landau levels \cite{CF}.

The picture of independent impurity potentials used in
Refs. \cite{convth,CF} is well justified in the case of the
standard experimental geometry \cite{exp}, where two Si 
delta-doped layers (DDLs) are situated on the outer sides
of the double quantum well. The DDLs not only serve as the
reservoirs supplying carriers but also introduce
disorder in the form of a smooth random potential (SRP) of the ionized
donors. Moreover, due to the 
efficient screening the carriers in each quantum well
experience only a SRP
created by the nearest DDL.

Instead, one can consider an alternative geometry
where a single DDL is located in the middle between the two
electron layers, so
that the SRPs in both layers are almost identical.
This setup gives one an opportunity to study a new type of coherent
effects in systems with spatially separated carriers.

In the present
paper  we investigate the influence of correlations between
the impurity potentials in different layers on the
transresistivity. We focus our attention onto the case
of a long characteristic time, $\tau_{g}$, at
which the carriers feel the difference between the SRPs in the
two layers ($\tau_g \gg \tau_{tr}$,
where $\tau_{tr}$ is the transport scattering
time in each layer).
We show that in this case the drag is strongly 
enhanced in comparison to the
non-correlated situation. This enhancement is due to
a possibility of a coherent motion of carriers propagating
in different layers and feeling nearly the same random potential. As
a result, the effective time of their interaction increases
considerably. This gives rise to the new behavior of the transresistivity
\begin{eqnarray} \rho^{corr}_{21} &\simeq&
-\frac{\pi^4}{24} \frac{\hbar}{e^2}
\frac{\ln( T \tau_g)}{(k_F d)^4 (\kappa l)^2}, \quad
\tau^{-1}_{g} \ll T \ll \tau^{-1}_{tr},
\label{rho1} \\
\rho^{corr}_{21}&\simeq& -
\frac{\pi^4}{6} \frac{\hbar}{e^2}
\frac{(T \tau_g)^2}{(k_F d)^4 (\kappa l)^2}, \quad T \ll
\tau^{-1}_{g},
\label{rho2}
\end{eqnarray}
Here, $l= v_F \tau_{tr}$ is the electron
mean free path,
$k_F\,(v_F) $ is the Fermi momentum (velocity),
$d$ is the interlayer distance (throughout this paper we assume
$l\gg d$), and $\kappa$ is the Thomas-Fermi
momentum. This term yields the dominant contribution to $\rho_{21}$
within the entire experimentally accessible temperature range,
provided that the system remains in the diffusive regime,
$T \ll \tau_{tr}^{-1}$. Below we specify the experimental
conditions necessary for the observations of the behavior
described by Eqs.\ (\ref{rho1}) and (\ref{rho2}), and
predict a suppression of these regimes
by a
weak
magnetic field.

In general, a diffusive motion of carriers
becomes coherent if there exist singularities in
the particle-hole and particle-particle
propagators  (``Diffusons'' and ``Cooperons'').
The equations for Diffusons in the double-layer system can be written
as follows 
\begin{equation}
{\cal D}^{ij}_{{\bf k} {\bf k'}}({\bf q},\omega) 
=
W^{ij}_{{\bf k} {\bf k'}} +
\sum_{{\bf k_1}}W^{ij}_{{\bf k} {\bf k_1}} G^{Ri}_{{\bf k_1}+{\bf q}}
(\epsilon+\omega)
G^{Aj}_{{\bf k_1}}(\epsilon){\cal D}^{ij}_{{\bf k_1} {\bf k'}}({\bf q},\omega).
\label{diffijeq}
\end{equation}
Here, the indices $i$,$j$ label the layers,
${\bf k}$ and ${\bf k^{\prime}}$
are the momenta of the incoming and the outgoing electrons,
$G^{R(A)i}_{{\bf k}}(\epsilon)=[\epsilon-\xi_{\bf k}\pm
i/2\tau^{ii}]^{-1}$
is the impurity averaged retarded (advanced) electron Green
function, and
$W^{ij}_{{\bf k} {\bf k'}}=\langle u^i u^j \rangle_{imp}$ are
the elastic electron scattering probabilities.
The values of $W^{ij}_{{\bf k} {\bf k'}}$ at $i\neq j$ differ
from zero due to correlations between the impurity
potentials $u^{i}$ in different layers. The total
and the transport scattering times are
defined, respectively, by the formulae
$$1/\tau^{ij}=\left<W^{ij}_{{\bf k} {\bf k'}}\right>_{{\bf k'}},
\quad
1/\tau^{ij}_{tr}=\left<W^{ij}_{{\bf k} {\bf k'}}
(1-\hat{\bf k}\hat{\bf k'})\right>_{{\bf k'}},$$
where the symbol
$\left<...\right>_{{\bf k}}=\nu_F\int^{2\pi}_0...d\varphi_k$
stands for the angle average over the Fermi surface, $\nu_F=m/2\pi$
is the single-spin density of states (assumed to be equal
in both layers), $m$ is the effective electron mass,
and $\hat{\bf k}={\bf k}/k$.

By solving the interlayer Diffuson equation with the use of
the formalism developed in Ref. \cite{woelfle}, we find
that it is the characteristic time 
\begin{equation}
\tau^{-1}_g=\frac{\tau^{21}-\tau}{\tau^2}
\label{taug}
\end{equation}
(where $1/ \tau = [1/ \tau^{11}+ 1/ \tau^{22}]/2$)
which determines a crossover between different regimes.
Namely, for $\tau^{ii}_{tr}\gtrsim \tau_g$ the interlayer
Diffusons do not form, and the system remains in the ballistic regime
at all temperatures, as far as the interlayer elastic scattering is concerned.
It is only at $\tau_g \gg \tau_{tr}^{ii}$ that 
the solution of Eq. (\ref{diffijeq}) 
\begin{equation}
{\cal D}^{21}_{{\bf k} {\bf k'}}({\bf q},\omega)\simeq
\frac{1}{2\pi\nu_F\tau^2}
\frac{\gamma_{\bf k}\gamma_{\bf k'}}{D q^2 - i\omega
+\tau_g^{-1}} + {\cal D}^{reg}_{{\bf k} {\bf k'}}
\label{diff12}
\end{equation} 
develops a quasi-Diffuson pole.
Here $D =v_F^2\tau_{tr}/2$ is the diffusion
coefficient and $\gamma_{\bf k}=1-i(\tau_{tr}-\tau){\bf q}{\bf
k}/m$.  
The regular part of the Diffuson ${\cal
D}^{reg}_{{\bf k} {\bf k'}}$ is given by the same
expression as in the single-layer case \cite{woelfle}.
We neglect the difference between the 
interlayer and intralayer scattering times by setting $\tau^{ij}=\tau$
and $\tau^{ij}_{tr}=\tau_{tr}$ everywhere except for the
"gap" $\tau^{-1}_g$ in
the first term of Eq.\ (\ref{diff12}).
For 
two identical 
SRPs one has $\tau_g=\infty$,
and the interlayer Diffuson coincides 
with the
intralayer one.

The same conclusions can be
reached about
the interlayer particle-particle propagator (Cooperon),
which
is given by Eq.\ (\ref{diff12}) with the substitution of
$\tau_{g}^{-1}$ by $\tau^{-1}_g+\tau_{\varphi}^{-1}$, where
$\tau_{\varphi}$ is the inelastic phase breaking time.

From Eq.\ (\ref{diff12}) one can see that at
$\tau_g \neq \infty$ there are no singularities in the
interlayer Diffusons and Cooperons. Nonetheless,
at frequencies
$\tau_g^{-1}\lesssim\omega \lesssim\tau_{tr}^{-1}$
and momenta $(\tau_{tr}/\tau_{g})^{1/2} \lesssim q l \lesssim
1$ the
motion of carriers in the two layers
is highly correlated.
%

The origin of the interlayer decoherence time $\tau_g$ can
be explained as follows. Consider two coherent electron waves
propagating in slightly different random
potentials ($u+ \delta u$ and
$u- \delta u$). After passing through the distance of order  the
SRP correlation length $a$ they acquire
a random phase
difference $\Delta\phi\sim (2\delta u)v_{F}^{-1} a $.
This leads to the electron`s phase diffusion  with the
diffusion coefficient $D_{ph}= (\Delta\phi)^2  v_F a^{-1}$,
and provides a complete loss of phase coherence over the time
$\tau_g \sim D_{ph}^{-1}$.



The new correlation effects for the
transconductivity are described by diagrams with two
electron loops (one current vertex per each), connected
not only by the interlayer
Coulomb interaction lines but also by the impurity lines
combining into the interlayer Diffusons and Cooperons.
We find that, to any order in the
interlayer interaction $V_{21}$, both the Diffuson and the
Cooperon diagrams for which the electron energy does not
change its sign at the current vertices give no
contribution to $\sigma_{21}$
at zero external frequency. In order to prove this statement we
extended the
method proposed in Ref.\
\cite{AALK} for treating the Hartree terms onto the double-layer
case. This general method
accounts for all
the diagrams with an arbitrary number of impurity lines (also in
the ballistic regime) and therefore remains valid in the presence
of the gap $\tau_{g}^{-1}$ in Diffusons and Cooperons.

Perturbative calculations
confirm the above statement.
To illustrate this, consider the
Diffuson diagrams which give rise to logarithmic conductivity
corrections in the single-layer case \cite{AA}.
Only two of such diagrams (see Fig.\ 1a) contain
the current vertices inserted in different electron loops and
hence are related to $\sigma_{21}$.
We found that the logarithmic terms in these diagrams
cancel against each other. To verify this in the case of
SRPs one has to take into account the ${\bf q}$-dependent
residues of the diffusons as well as their regular parts
(cf.\  Ref.\ \cite{woelfle}). Therefore, in the
double-layer system only the intralayer conductivities
$\sigma_{ii}$ acquire the logarithmic Hartree corrections
due to the presence of the second layer.

Thus the only
diagrams contributing to $\sigma_{21}$ are
those with the energy sign
changing at the current vertices. It can be checked that to
any order in $V_{21}$
no Diffuson diagrams of this kind survive in the DC limit.
On the contrary, the Cooperon diagrams do, and the first
nontrivial contribution to $\sigma_{21}$ arises in the
second order in the interaction $V_{21}$ and involves
three Cooperons. It is, however,
more instructive to sum up the entire interlayer
Coulomb-Cooperon ladder (see Fig.\ 1b). Taking into account
the smooth character of the donors' potentials, one arrives at
the two diagrams depicted in Fig.\ 1c. After the summation
over electron frequencies and momenta, the contribution of
these diagrams to the transconductivity takes the form

\begin{equation}
\sigma_{21}^{corr} 
=
\frac{4 e^2}{\pi \hbar T} \int \frac{D(dq)}{D q^2
+ \tau_{g}^{-1}+\tau_{\varphi}^{-1}}
\int_{0}^{\infty} \! \!
\frac{d \omega}{\sinh^2{\frac{\omega}{2T}}} \,
{\rm Im} \Psi_c ({\bf q}, \omega) \, {\rm Im} \Lambda_c ({\bf q}.
\omega),
\label{MT}
\end{equation}
The quantities $\Psi_c$ and $\Lambda_c$ are given by
\begin{eqnarray}
\Psi_c ({\bf q}, \omega) &=& \psi \left( \frac{D q^2 - i \omega +
\tau_{g}^{-1} + \tau_{\varphi}^{-1} }{4\pi T} + \frac{1}{2} \right),
\nonumber \\
\Lambda_c ({\bf q}, \omega) &=& 2 \, \left[ \ln \frac{\varepsilon_0}{T}
+ \lambda_{21}^{-1} - \Psi_c ({\bf q}, \omega) + \psi (1/2) \right]^{-1},
\nonumber
\end{eqnarray}
where $\psi$ is the digamma function, $\varepsilon_0\propto
\varepsilon_F$ is the upper energy cutoff, and $\lambda_{21}$ is
the effective interaction constant. The above equations are
valid with a logarithmic accuracy, since we omitted the terms
originating from the regular parts of the Cooperons.
The quantity $\lambda_{21}$ is defined as
$\lambda_{21}=(4\pi^2 \nu_F)^{-1}
\left<V_{21}({\bf p}-{\bf p'})\right>_{{\bf p},{\bf p'}}$,
with the screened interlayer 
Coulomb interaction being of the form
$$ V_{21}({\bf p})=\frac{1}{2\nu_F}\left[
\frac{\kappa(1+e^{-pd})}{p+\kappa(1+e^{-pd})}-
\frac{\kappa(1-e^{-pd})}{p+\kappa(1-e^{-pd})}\right].$$
Assuming that the screening is
strong enough, $\kappa d \gg 1$, and that the interlayer Coulomb
potential is sufficiently smooth, $k_F d \gg 1$,
one finds $\lambda_{21}\simeq \pi (4 k_F d \kappa d)^{-1}$.

Evaluation of the integrals in Eq. (\ref{MT}) yields
\begin{equation}
\rho^{corr}_{21}\simeq -\frac{2\pi^2}{3}
\frac{\hbar}{e^2} \frac{1}{(k_F l)^2
[\lambda^{-1}_{21} + \ln(\varepsilon_0 /T)]^2}
\ln\frac{T\tau_{\varphi}\tau_{g}}
{\tau_{\varphi}+\tau_{g}}
\label{rhototal1}
\end{equation}
at $\tau^{-1}_{g} \ll T \ll \tau^{-1}_{tr}$
(when the domain of ${\bf q}$-integration
is effectively limited by $D q^2 \lesssim T$), and
\begin{equation}
\rho^{corr}_{21}\simeq -\frac{8\pi^2}{3}
\frac{\hbar}{e^2} \frac{(T\tau_g)^2}{(k_F l)^2
[\lambda^{-1}_{21} + \ln(\varepsilon_0 \tau_g)]^2}
\label{rhototal2}
\end{equation}
at lower temperatures.
These equations constitute our main result.
Under realistic experimental conditions 
the value of 
$\lambda_{21}^{-1}$ is sufficiently large for one to neglect the logarithmic
terms
in the denominators of Eqs.\ (\ref{rhototal1}) and
(\ref{rhototal2}).
Also, since the interlayer decoherence time $\tau_g$ is
temperature independent, the argument of the logarithmic
function in the numerator of
Eq.\ (\ref{rhototal1}) is linear in temperature provided that
$\tau_g \ll \tau_{\varphi}$.  
Then Eqs.\ (\ref{rhototal1}) and (\ref{rhototal2}) reduce to
Eqs.\ (\ref{rho1}) and (\ref{rho2}), respectively.

Now let us compare 
these equations with the results of the 
standard theory \cite{convth}:
\begin{equation}
\rho^{conv}_{21}=\frac{\hbar}{e^2}\frac{\pi^2 \zeta(3)}{16}
\frac{1}{(k_Fd)^2(\kappa d)^2}
\left(\frac{T}{\varepsilon_F}\right)^2.
\label{roconv}
\end{equation}
We see that at $T\ll \tau_{tr}^{-1}$ our result
exceeds the conventional one:
in the interval $\tau_{g}^{-1} \ll T \ll \tau_{tr}^{-1}$ the
correlation effects lead to the smoother T-dependence,
while at $T \ll \tau^{-1}_g$ the prefactor of the $T^2$-dependence 
is $(\tau_g/\tau_{tr})^2$ times greater in our case.

Two analogies should be mentioned. Firstly, Eq.\ ({\ref{MT}})
resembles the Maki-Thompson correction to the conductivity
of a single-layer system \cite{MT}. However, in that case
the corresponding processes yield a small correction to
the Drude term while in the double-layer system
they determine the leading contribution to $\sigma_{21}$.
Also, in our situation there exists the
temperature-independent quantity $\tau_g$
resulting in a new behavior at $\tau_g
\ll \tau_{\varphi}$ and $T \ll \tau_g^{-1}$. We note that
this quantity plays a role similar to
that of
the magnetic field.
On the other hand,  a perpendicular magnetic
field leads to a suppression of the transresistivity, since one has to replace
$\tau_{g}^{-1}$ by $\tau_{H}^{-1} = 4 D e H / (\hbar c)$ in
Eq.\ (\ref{rho1}) at $\tau_{g}^{-1} \ll \tau_{H}^{-1} \ll
T$ and in Eq.\ (\ref{rho2}) at $T, \tau_{g}^{-1} \ll
\tau_{H}^{-1}$. This effect of the magnetic
field could provide  a test for the theory.

Secondly, it can be seen from Eq.\ (\ref{MT}) that the nontrivial
contribution to $\sigma_{21}$ arises in the
second order in $\lambda_{21}$ in which case one should
replace ${\rm Im}\Lambda_c$ by  $2\lambda_{21}^{2}\, {\rm
Im}\Psi_c $.
The resulting expression
looks quite similar to the usual one obtained for
the uncorrelated impurity potentials \cite{convth}, 
since they both describe the second-order Coulomb interaction
processes in terms of 
 some effective bosonic modes. The
difference is in the physical meaning of these bosonic modes,
which are the (energy-integrated) interlayer
Cooperons versus the interlayer plasmons in the case considered
in this paper and
in the conventional situation, respectively. Also, there is
an extra $q^2$ factor in the latter case providing
the rapid decay of $\rho^{conv}_{21}(T)$.

In the above consideration we neglected
the Aslamazov-Larkin --- type diagrams \cite{AL} which
contain two $\Lambda_c$ ladders, since being proportional to $T^2
\lambda^{2}_{21}$, these terms appear to be of the order of the ordinary
contribution given by Eq.\ (\ref{roconv}).

Now we discuss the experimental conditions
under which the above theory applies. For the standart geometry we have found
that at $k_F a> 1$ and $k_F d > 1$ (here $a$ is the
distance from DDL to the nearest quantum well) the
condition $\tau_g\gg\tau_{tr}$ can never be satisfied as long as
$\kappa d > 1$.  At $\kappa d< 1$ and $\kappa a> 1$ 
it requires $2(\kappa d)^2(k_F a)^2 < 1$, 
which can only be possible at 
very  small interlayer distances \cite{note}.

The situation is different for the suggested geometry with a single DDL
located {\it between} the two quantum wells. Introducing a finite
width of the DDL  $\delta$ we find that at $k_F d>
1$ and $(2\delta/d)^2<(\kappa d)^{-1}$ the condition
$\tau_g\gg\tau_{tr}$ can be rewritten as
\begin{equation}
2(k_F\delta)^2\ll {\rm min}[1,\kappa d]
\label{cond}
\end{equation}
For $\kappa\sim 0.02$ {\AA}$^{-1}$, $k_F\sim 0.015$ {\AA}$^{-1}$,
$\delta\sim 10$ {\AA}, and $d \sim 400$ {\AA}
the above criteria are fulfilled, and there exists the regime of
temperatures 
described by Eq.\ (\ref{rho1}). Note that it might be easier to
observe this regime in dirty samples (yet with 
$l\gg d$). For $l \sim 5000$ \AA (which implies
$\tau_{tr}^{-1} \sim 4 $K and $\tau_g^{-1}\sim 0.2 $K)
Eq.\ (\ref{rho1}) yields $\rho_{21}$ of the order of a
few m$\Omega$s
within the entire
temperature range $\tau_{tr}^{-1} \gtrsim T  \gtrsim
\tau_g^{-1}$,
whereas the conventional theory would predict a rapid decay
of the transresistivity from $\rho_{21} \sim 1$ m$\Omega$ at $T
\sim \tau_{tr}^{-1}$ to $\rho_{21} \sim 3 \mu \Omega$ at
$T \sim \tau_{g}^{-1}$.

In conclusion, we investigated the influence of correlations
between impurity potentials in different layers of
a double-layer electron system on the
Coulomb drag effect. We found that for  
correlated potentials the low-temperature drag
is substantially enhanced compared  to the standard
( uncorrelated) situation, whereas a
weak magnetic field suppresses the effect.

We gratefully acknowledge useful discussions with
A.P.\ Dmitriev,
M.I.\ Dyakonov,
A.L.\ Efros,
D.\ Golosov,
B.Yu-K.\ Hu,
V.Yu.\ Kachorovskii,
A.V.\ Khaetskii,
A.V.\ Subashiev, and
I.\ Ussishkin.
This work has been supported by INTAS Grant No.\ 97-1342,
and in part by RFBR (IG and AY).
IG and AY are also thankful to NORDITA for
financial support and hospitality during their stay in
Denmark.



{\it Figure caption}.
Diffuson diagrams for transconductivity (a), interlayer
Coulomb-Cooperon ladder (b), and two Cooperon diagrams
which yield the main contribution to
$\sigma_{21}$ in the considered case (c).
Solid lines stand for electron Green
functions,
wavy lines, and single dashed lines denote
the  screened Coulomb interaction and the
impurity potential, respectively. Double dashed lines
represent the Diffusons and Cooperons, and the
numbers label the layers.

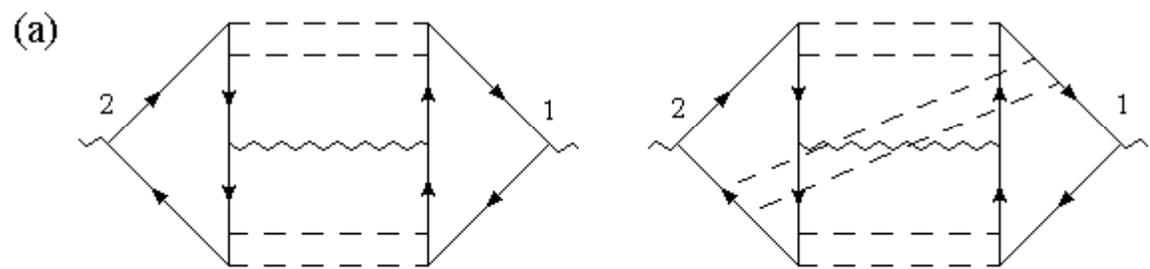
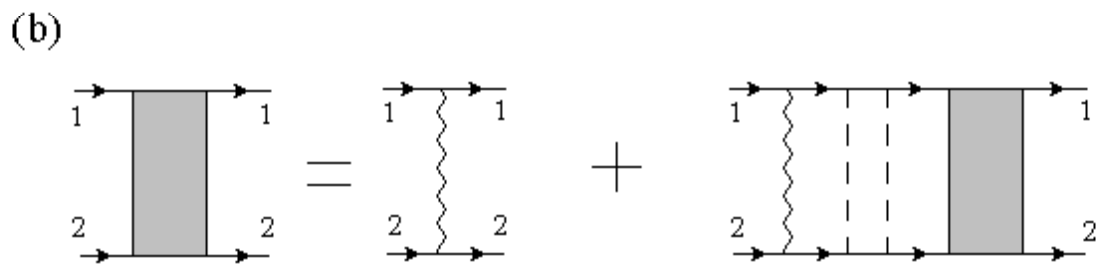
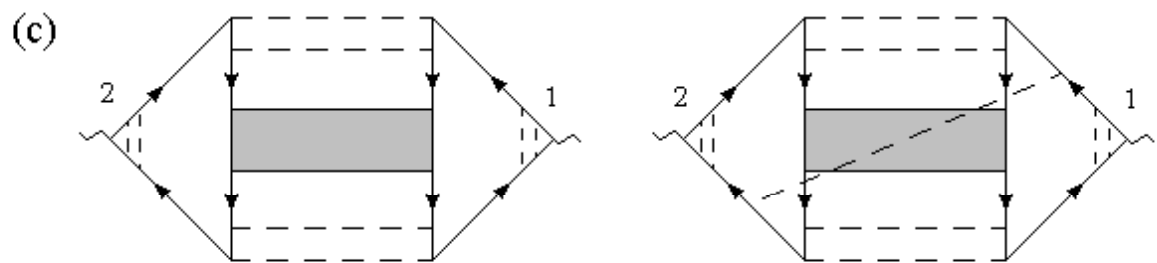
\end{document}